\documentclass{article}

\usepackage{arxiv}

\usepackage[utf8]{inputenc} 
\usepackage[T1]{fontenc}    
\usepackage{hyperref}       
\usepackage{url}            
\usepackage{booktabs}       
\usepackage{amsfonts}       
\usepackage{nicefrac}       
\usepackage{microtype}      
\usepackage{lipsum}
\usepackage{graphicx}
\usepackage{amsmath}
\graphicspath{ {./images/} }

\title{Classification of lung cancer subtypes on CT images with synthetic pathological priors}

\author{
Wentao Zhu \\
  Research Center for Healthcare Data Science\\
  Zhejiang Lab\\
  Hangzhou 311121, China \\
   \And
 Yuan Jin \\
  Research Center for Healthcare Data Science\\
  Zhejiang Lab\\
  Hangzhou 311121, China \\
  \And
 Gege Ma \\
  Research Center for Healthcare Data Science\\
  Zhejiang Lab\\
  Hangzhou 311121, China \\
  \And
 Geng Chen \\
  School of Computer Science and Engineering\\
  Northwestern Polytechnical University\\
  Xi'an, Shaanxi 710072, China \\
  \And
 Jan Egger \\
  Institute of Computer Graphics and Vision\\
  Graz University of Technology\\
  8010 Graz, Austria \\
  \And
 Shaoting Zhang \\
  Shanghai Artificial Intelligence Laboratory\\
  Shanghai 200120, China\\
  \And
 Dimitris N. Metaxas \\
  Department of Computer Science\\
  Rutgers University\\
  Piscataway, NJ 08854, USA \\
}

\begin{document}
\maketitle
\begin{abstract}
The accurate diagnosis on pathological subtypes for lung cancer is of significant importance for the follow-up treatments and prognosis managements. In this paper, we propose self-generating hybrid feature network (SGHF-Net) for accurately classifying lung cancer subtypes on computed tomography (CT) images. Inspired by studies stating that cross-scale associations exist in the image patterns between the same case's CT images and its pathological images, we innovatively developed a pathological feature synthetic module (PFSM), which quantitatively maps cross-modality associations through deep neural networks, to derive the ``gold standard” information contained in the corresponding pathological images from CT images. Additionally, we designed a radiological feature extraction module (RFEM) to directly acquire CT image information and integrated it with the pathological priors under an effective feature fusion framework, enabling the entire classification model to generate more indicative and specific pathologically related features and eventually output more accurate predictions. The superiority of the proposed model lies in its ability to self-generate hybrid features that contain multi-modality image information based on a single-modality input. To evaluate the effectiveness, adaptability, and generalization ability of our model, we performed extensive experiments on a large-scale multi-center dataset (i.e., 829 cases from three hospitals) to compare our model and a series of state-of-the-art (SOTA) classification models. The experimental results demonstrated the superiority of our model for lung cancer subtypes classification with significant accuracy improvements in terms of accuracy (ACC), area under the curve (AUC), and F1 score.
\end{abstract}


\section{Introduction}
\label{sec:introduction}
\begin{figure*}[!t]
     \centering
     \begin{tabular}{cccc} 
        \large{LUAD} &  \includegraphics[width=0.35\textwidth]{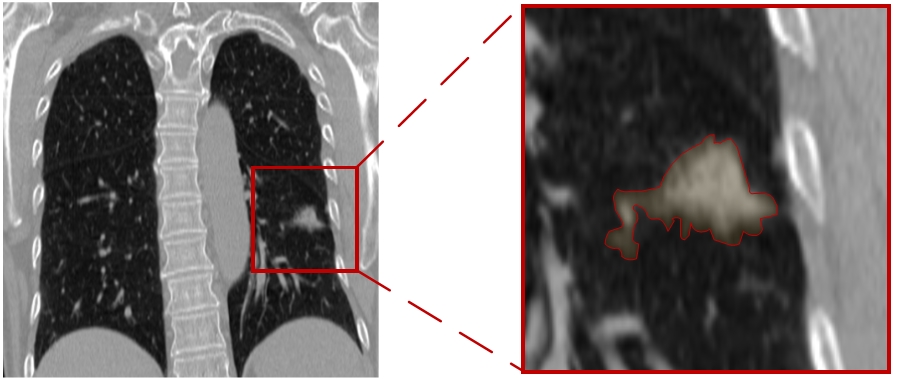}  & & \includegraphics[width=0.35\textwidth]{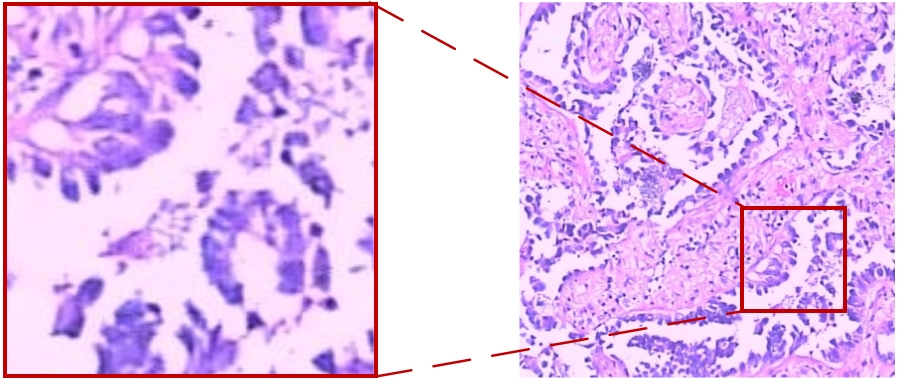} \\
                  & (a) & & (b)  \\
         \large{LUSC} &  \includegraphics[width=0.35\textwidth]{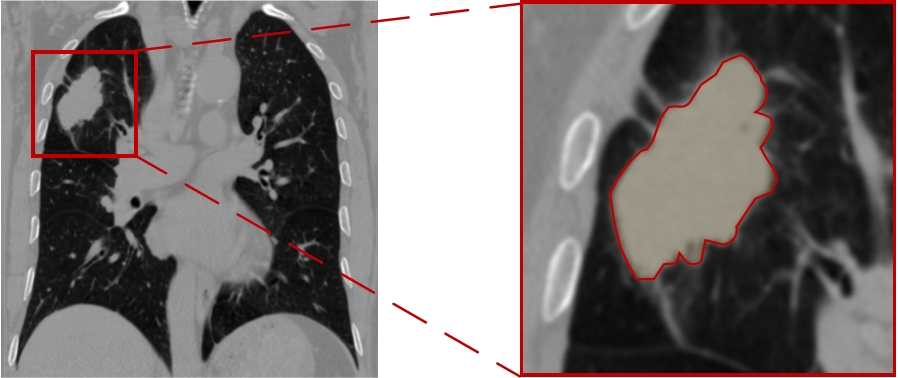} & & \includegraphics[width=0.35\textwidth]{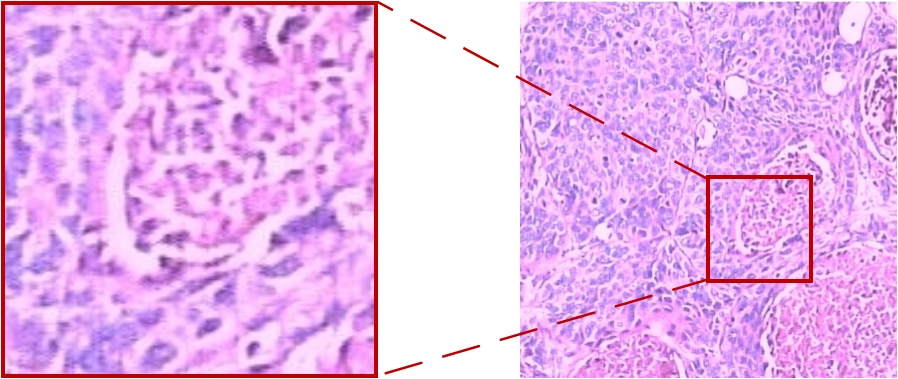} \\
                  & (c) & & (d) \\
      \end{tabular}
\caption{The detailed image patterns in the paired CT images and pathological images of LUAD case (a) $\&$ (b), and LUSC case (c) $\&$ (d).}
    \label{fig:matchedCT_WSI}
\end{figure*}
Lung cancer is one of the most prevalent malignant tumors and is the leading cause of cancer-related mortality globally \cite{ferlay2019estimating}. It has been estimated that 2.2 million new cases of lung cancer and 1.8 million deaths occurred worldwide due to this disease in 2020 \cite{sung2021global}. Lung cancer is a complex and diverse disease. According to the origins of tissues and the biologic behaviors of tumors, lung cancer can be classified into small-cell lung carcinoma (SCLC) and non-small-cell lung carcinoma (NSCLC) \cite{minna2002focus}. Approximately 85\% of lung cancers are NSCLC, of which lung adenocarcinoma (LUAD) and lung squamous cell carcinoma (LUSC) are the histological subtypes with the highest clinical incidence rates \cite{nicholson20222021,travis2020lung}. Studies have shown that the vast diversity between LUAD and LUSC can be revealed at the molecular, pathological, and clinical levels \cite{wang2020comparison,relli2019abandoning}. Consistent with their diversity, the responses of them to the same therapeutic strategy may also be distinct \cite{relli2019abandoning,herbst2018biology}. In \cite{forde2018neoadjuvant}, the activity of immune checkpoint inhibitors was found to be different in LUAD than that in LUSC. In addition, Scagliotti et al. reported varying outcomes for LUAD and LUSC patients treated with the chemotherapy drug pemetrexed \cite{scagliotti2011treatment}. More recently, targeted anti-vascular endothelial growth factor (VEGF) bevacizumab therapy was found to be effective in treating patients with LUAD, whereas it was listed as a contraindication to LUSC \cite{kalemkerian2018molecular}. Therefore, as lung cancer is a disease with rapid development and a poor prognosis, performing accurate pathological subtypes diagnosis of lung cancer at an early stage is critical for effective and personalized therapeutic managements. 

A variety of clinical techniques have been developed to assist physicians in diagnosing lung cancer, such as chest radiography, computed tomography (CT), bronchoscopy, pathological examination and so forth \cite{prabhakar2018current}. Among these modalities, fast imaging and non-invasive CT scan has become the most commonly used method for early cancer detection and diagnosis \cite{collins2007lung}. By offering the three-dimensional anatomical image, CT is informative in terms of revealing the sizes, shapes, positions, metastasis statuses and heterogeneity of tumors. For certain pathological subtypes of lung cancer, several radiological manifestations can be utilized as diagnostic indicators for preliminary assessments \cite{jiang2014thin,cohen2016lung,yue2018ct,gharraf2020role}. However, as the interpretations of radiographic results depend heavily on clinical experience, diagnostic opinions may vary among physicians. Moreover, tumors at the early stage may lack typical clinical manifestations, which makes it difficult to detect subtle pathological changes through conventional visual assessments. As a result, a highly accurate computer-aided CT analysis system for diagnosing lung cancer subtypes is in high demand.

Deep learning (DL), as an automatic quantification method, is regarded as an effective and promising approach for addressing the aforementioned problems \cite{razzak2018deep}. By automatically extracting and analyzing high-throughput features from radiographic images with specific end-to-end deep neural networks, DL enables the quantitative identification of the image variations exhibited by different lesions and subsequently yields diagnostic and prediction accuracy improvements. In recent years, milestones have been achieved with DL in the field of automatic classifications on CT images through various proposed convolutional neural networks (CNNs) and learning strategies \cite{xie2018knowledge,lakshmanaprabu2019optimal,amyar2020multi, wang2020classification}. These approaches hold advantages over visual assessments by providing physicians with more rapid and accurate diagnostic assistance to some extent. However, for the complex task of automatically classifying cancer subtypes on CT images, the classification accuracies and robustness levels of existing models may still not be satisfactory. Likewise, the limitations of these classification models can be attributed to the atypical radiological manifestations of certain cases. In addition, the original CT images usually carry a great deal of redundant information, which is also a non-negligible obstacle that prevents DL algorithms from reaching satisfactory precision rates.

\begin{figure*}[!t]
	\centering
	\includegraphics[width=\textwidth]{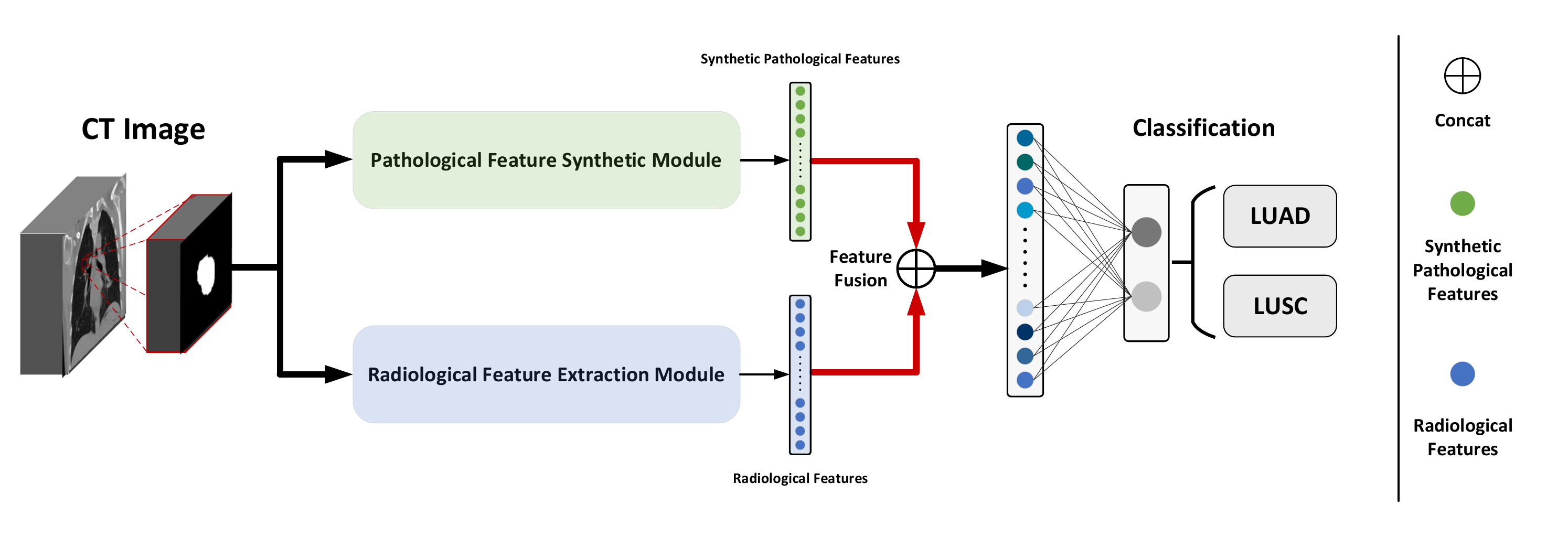}
	\caption{Pipeline of the proposed novel CT-based classification model, SGHF-Net: $P=C(F(f_{\text{p}},f_{\text{r}}))$, where $C(\cdot)$ denotes the classification component; $F(\cdot)$ denotes the fusion module; $f_{\text{p}}$ and $f_{\text{r}}$ are the pathological and radiological features, respectively.}
	\label{fig:structure}
\end{figure*}
In clinical practice, to obtain the most accurate diagnoses of cancer subtypes, pathological examination is highly recommended as a further complementary test. This modality is regarded as the ``gold standard" in cancer diagnosis since the pathological images obtained during the examination contain information of cell morphology, cell differentiation degree, cell density and other information. Hence, providing the information contained in pathological images as additional knowledge for a conventional radiological feature (RF)-based model may promote it to output a more accurate prediction. However, it is worth noting that pathological examinations are invasive, as they require tissue specimens to be obtained through needle biopsy or surgical resection \cite{prabhakar2018current}. In some cases, the physical conditions of patients or the potential risks of complications may limit the possibility of utilizing CT-guided needle biopsy \cite{wu2011complications,haramati1991complications}. Thus, pathological examination may not always be readily available in the early diagnosis phase.

In nature, the CT and pathological images of the same lesion are the visual expressions of pathological tissues at different spatial scales and in different resolutions. Recently, there have been increasing interests in identifying the cross-scale associations between pathological images and radiographic images through high-level pathological and radiological feature approaches. \cite{ganeshan2013non} found that the radiological features that capture the heterogeneity of NSCLC from contrast-enhanced CT images have correlations with certain histopathologic markers generated due to hypoxia and angiogenesis. \cite{lederlin2013correlation} reported distinct associations between the radiological and histomorphological patterns of LUAD, where the tumor margin configurations and solidity glass opacity levels on CT images correspond to certain cell growth patterns. In \cite{alvarez2020identifying}, Alvarez-Jimenez further identified the cross-scale associations between the radiological and pathological features of NSCLC by showing the relationships between CT intensity values and matched cell density statistics. Moreover, Khorrami confirmed the effectiveness of using the radiological features associated with lymphocyte distributions to predict the therapy outcomes and survival rates of NSCLC patients \cite{khorrami2020changes}. Based on these cross-modality correlation findings, optimal strategies can be developed to obtain the underlying pathological information of tumors from CT images.   

In this study, we propose a novel self-generating hybrid feature network (SGHF-Net) for accurately classifying lung cancer subtypes on CT images. Inspired by studies on the cross-modality associations between CT images and pathological images, we propose to exploit these correlations with DL techniques to acquire the gold-standard pathological image features from CT images. More importantly, we develop an effective feature fusion framework to integrate these synthetic pathological features (SPFs) into a benchmark RF-based model, guiding the entire classification model to be more inclined to extract the pathologically relevant features from CT inputs. The main property of our model is that it takes paired CT and pathological images as training data while only requires the CT images in the subsequent validation periods. In this way, our model enables the self-generation of multi-modality hybrid features while relying on a single image source (i.e., CT images) in clinical applications. Such an approach not only breaks the information deficiency limitations of a single diagnostic modality but also compensates for the absence of synergy between different modalities in the traditional one-way diagnostic process. Moreover, the proposed pathological priors guided strategy can be adopted by many state-of-the-art (SOTA) classification networks without incurring extra costs. To summarize, the key contributions and novelties of this work are listed as follows.  
\begin{itemize}
  \item We design a pathological feature synthetic module (PFSM), which quantitatively maps the cross-scale associations between dual medical imaging modalities, to generate the high-level deep features of pathological images from CT images.   
  \item We design a radiological feature extraction module (RFEM) to directly acquire the radiographic information contained in CT images and integrate it with the PFSM under an effective feature fusion framework, forming more indicative and robust hybrid features for cancer subtypes classification. 
  \item We build a large-scale multi-center dataset with data from three different tertiary hospitals. Through a series of experiments, the results all confirm the superiority of our approach for lung cancer subtypes classification.
\end{itemize}
\begin{figure*}[!t]
	\centering
	\includegraphics[width=\textwidth]{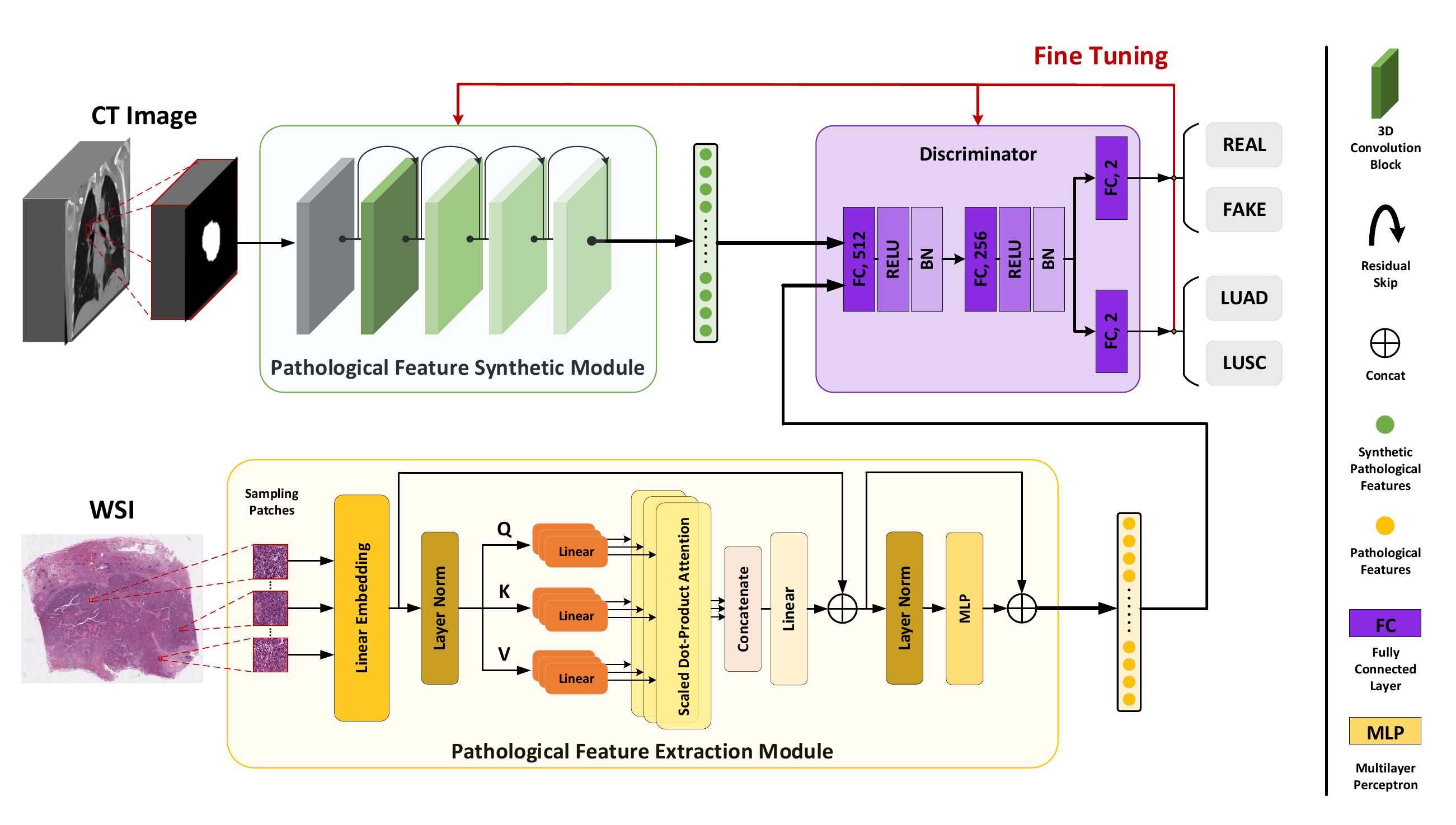}
	\caption{The training procedure of the \textbf{PFSM} (notably, the pathological images, as the gold-standard reference images, are only required during the training process of the PFSM.)}
	\label{fig:S2}
\end{figure*}

\section{Method}
\label{sec:Methods}
In this section, the proposed novel lung cancer subtypes classification model, SGHF-Net, is introduced. The inherent challenge of applying DL approaches in complex CT-based classification scenarios is the hard-to-achieve high accuracy. Restricted by the finiteness of valid manifestations and the interference caused by redundant information, it is highly challenging to achieve satisfactory performance with a model that relies on features from a single image modality (i.e., CT images). Therefore, as depicted in Fig. \ref{fig:structure}, we designed two key components for the proposed model, namely, the pathological feature synthetic module (PFSM) and the radiological feature extraction module (RFEM), to utilise CT images to self-generate hybrid features containing information from two image modalities. In the following subsections, the functions and training details of each module are elaborated.

\subsection{Pathological Feature Synthetic Module}
The PFSM was designed to quantitatively map the correlations between the dual medical image modalities, so as to derive the ``gold standard" information contained in the corresponding pathological images from CT images. Throughout the training phase, the PFSM was trained firstly with paired pathological and CT images, and then the well-trained PFSM was integrated into the benchmark RF-based model, working as pathological priors to co-supervise the RFEM parameters training process with the ground truth. Fig. \ref{fig:S2} illustrates the training procedure of the PFSM, which involves the procedures of extracting the pathological features from the pathological image patches and synthesizing the corresponding pathological features with CT images. Notably, the pathological images involved in this work are whole-slide images (WSIs), which are the digitized versions of glass slides that have been processed by a dedicated slide scanner \cite{al2012whole}.
\label{subsec:PFE}
\subsubsection{Pathological Feature Extraction from WSIs}
The high-level pathological features extracted from pathological images are the crucial cornerstones during the PFSM training process. Here, a subtype classification model with pathological images as inputs was pretrained, and once the training process was completed, its feature extraction blocks were taken out separately to extract the most representative high-level features for the subsequent PFSM training step. Notably, the desired high-level features in this work could be represented as a $512\times1$ feature vector. We adopted the well-known Vision Transformer (ViT) \cite{Dosovitskiy2020vit} as the pathology-based classification model and carried out the following operations. 

First, the original pathological image obtained after preprocessing was cropped into patches with identical sizes of 560$\times$560. Then, the cropped patches with more than 80\% cancer coverage were selected as the model inputs. To make the model jointly attend the information contained in these patches from different sub-spaces, we applied the multi-head self-attention mechanism in the network architecture,
\begin{equation}
\operatorname{MultiHead}(Q, K, V)=\operatorname{Concat}(head_1,...,head_h)W^O
\label{eq:Multihead}
\end{equation}
where
\begin{equation}
head_t=\operatorname{Attention}(QW^Q_t, KW^K_t, VW^V_t)
\label{eq:head}
\end{equation}
\begin{equation}
\operatorname{Attention}(Q_t, K_t, V_t)=\operatorname{softmax}\left(\frac{Q_t K^T_t}{\sqrt{d_k}}\right) V_t
\label{eq:Attention}
\end{equation}

where $Q$, $K$, and $V$ refer to the query, key, and value parameters of pathological patches, respectively, $t$ is the $t^{th}$ head, and $W^Q_t \in \mathbb{R}^{d_{model} \times d_k}, W^K_t \in \mathbb{R}^{d_{model} \times d_k}, W^V_t \in \mathbb{R}^{d_{model} \times d_v}$ and $W^O \in \mathbb{R}^{hd_{v} \times d_{model}}$ are weighted parameters. Here, $h=12$, and $d_{model}/h=d_k=d_v=64$. Following the architecture of the Transformer network and relying on self-attention, the model then outputted the weighted sum of the values of all input pathological patches.

During training, the whole dataset was divided into a training dataset and a testing dataset on a pro-rata basis (i.e., 80\%: 20\%). Following five-fold cross-validation, the patches belonging to different subjects and their corresponding pathological labels were crossly applied for model training. To update the network parameters, we applied the cross-entropy loss, as defined in Eq. \eqref{eq:CEL}, as the loss function:
\begin{equation}
\begin{aligned}
\operatorname {\mathcal{L}_{\text{PI}}} =-\frac{1}{N} \sum_{i} [y_{i} \cdot \log \left(p_{i}\right)+(1-y_{i}) \cdot \log \left(1-p_{i}\right)],
\end{aligned}
\label{eq:CEL}
\end{equation}
where ${\mathcal{L}_{\text{PI}}}$ refers to the loss, $N$ refers to the total number of training samples, and $i$ refers to the $i^{th}$ sample. $y_{i}$ is the label of the $i^{th}$ sample, and it is a sign function, which equals one if the label is the same as LUSC and zero otherwise. $p_{i}$ is the probability that the prediction of the $i^{th}$ sample belongs to the LUSC class. \\
\subsubsection{Pathological Feature Synthesis with CT Images}
\label{subsec:PFG}
Synthesizing the pathological features with CT images was a critical step of this work, which was achieved with a conditional generative adversarial network (CGAN). First introduced by Goodfellow et al. in 2014 \cite{goodfellow2020generative}, the generative adversarial network (GAN) has produced remarkable outcomes in a variety of applications \cite{aggarwal2021generative,yi2019generative}, including image super-resolution reconstruction \cite{mahapatra2019image}, image classification \cite{teramoto2020deep}, texture transferring \cite{li2016precomputed}, cross-modality synthesis \cite{nie2017medical}, and etc. The GAN, whose framework design was inspired by the two-player minimax game, is a special type of generative models. It is composed of two neural networks, a generator $\mathcal{G}$ and a discriminator $\mathcal{D}$. The generator aims to capture the distribution of an input dataset, while the discriminator seeks to determine the probability that a sample is from the real data distribution rather than $\mathcal{G}$ \cite{goodfellow2020generative}. By simultaneously training both networks in an adversarial manner, the generator and discriminator eventually reach a dynamic equilibrium, where the data distribution generated by $\mathcal{G}$ is close to the real data distribution, and the probability predicted by $\mathcal{D}$ equals 0.5. Due to the successful applications of GAN in many fields, much effort has been dedicated to improving the performance of GAN. For instance, plenty of studies have suggested that adding additional tasks to a GAN is beneficial for improving the performance achieved on the original REAL/FAKE discrimination task \cite{odena2017conditional,odena2016semi}. Inspired by these studies, we utilized $\mathcal{G}$ to generate high-level feature vectors of CT images with a specific class label (denoted as $c$). Furthermore, $\mathcal{D}$ contained two functions, including (i) discriminating whether a feature vector was generated with the pathological images or the CT images, and (ii) predicting the class label for this feature vector. The ultimate goal of our CGAN training was for the discriminator to fail to identify which imaging modality a feature vector was generated from while enabling to provide the correct subtype classification for this feature vector. 

The networks of our proposed $\mathcal{G}$ and $\mathcal{D}$ were implemented with multiple convolutional blocks and fully-connected (FC) layers. Specifically, ResNet and DenseNet were evaluated in comparison as the backbone of the $\mathcal{G}$ network. The discriminator $\mathcal{D}$ was composed of three FC layers, where each of the first two FC layers was followed by a rectified linear unit (ReLU) activation layer and a batch normalization (BN) layer. Besides, there were two output predictions, i.e., REAL/FAKE (task one) and LUAD/LUSC (task two). During the training process, the network parameters of $\mathcal{G}$ and $\mathcal{D}$ were defined as
\begin{equation}
\left\{\begin{array}{lr}
\theta_{\mathcal{G}}=W_{\mathcal{G}};b_{\mathcal{G}}     \\
\theta_{\mathcal{D}}=W_{\mathcal{D}};b_{\mathcal{D}} 
\end{array}\right.
\end{equation}
where $W_{\cdot}$ and $b_{\cdot}$ refer to the weights and bias of the network. $\theta_{\mathcal{G}}$ and $\theta_{\mathcal{D}}$ were updated through iterative optimization by solving a min-max problem. The input of $\mathcal{G}$ (i.e., the CT image with a corresponding class label $c$) was denoted as $z$, and the output of $\mathcal{G}$ (i.e., the $512\times1$ feature vector generated from CT image) was denoted as $x_{\mathcal{G}}$. We defined the corresponding probability distributions of $z$, $c$ and $x_{g}$ as $p_{z}$, $p_{c}$, and $p_{g}$. The nonlinear mapping from $z$ to $x_{\mathcal{G}}$ can be defined as 
\begin{equation}
x_{\mathcal{G}}=\mathcal{G}(z;\theta_{\mathcal{G}})
\end{equation}
In practice, we expected to maximize the similarity between $x_{\mathcal{G}}$ (fake sample) and the target vector $x_{r}$ (real sample), which was a $512\times1$ feature vector extracted from the pathological image. For $\mathcal{D}$, its input $x$ was either the generated sample $x_{\mathcal{G}}$ or the real sample $x_{r}$. We then formulated the learning process of $\mathcal{D}$ as
\begin{equation}
(y_{1},y_{2})=\mathcal{D}(x;\theta_{\mathcal{D}})
\end{equation}
where $y_{1} \rightarrow(0,1)$, and $y_{2} \rightarrow (0,1)$ are the two probability distributions for task one and task two, respectively. $y_{1}$ reflects the probability of $\mathcal{D}$'s input coming from the real sample group. It equals one if the input $x$ comes from $x_{r}$, and zero if $x$ comes from $x_{g}$. $y_{2}$ reflects the probability of different cancer subtypes. It equals one if the label is LUSC and zero if the label is LUAD.

During the training process of the CGAN, the overall objective function, which was designed to maximize the log likelihood of the correct source and the correct class label, was composed of two parts, $(\mathcal{L}_{\mathcal{D}1},\mathcal{L}_{\mathcal{G}1})$ and $(\mathcal{L}_{\mathcal{D}2},\mathcal{L}_{\mathcal{G}2})$. For task one, the mathematical expressions of the min-max problem for $\mathcal{D}$ and $\mathcal{G}$ can be defined as: 
\begin{equation}
\begin{aligned}
\mathcal{L}_{\mathcal{D}1} =&\max _{\mathcal{D}} \mathbb{E}_{x_{r} \sim p_{r}(x)}\left[\log \mathcal{D}\left(x_{r}\right)\right]\\&+\mathbb{E}_{x_{\mathcal{G}} \sim p_{g}(x)}\left[\log \left(1-\mathcal{D}\left(x_{\mathcal{G}}\right)\right)\right]
\end{aligned}
\label{eq:lossD1}
\end{equation}
\begin{equation}
\begin{aligned}
\mathcal{L}_{\mathcal{G}1} =\min _{\mathcal{G}} \mathbb{E}_{x_{\mathcal{G}} \sim p_{g}(x)}\left[\log \left(1-\mathcal{D}\left(x_{\mathcal{G}}\right)\right)\right]
\end{aligned}
\label{eq:lossG1}
\end{equation}
where $p_{r}$ is the probability of real data distribution. Eq. \eqref{eq:lossD1} and Eq. \eqref{eq:lossG1} train the generator to minimize the difference between the generated sample and the real sample, eventually encouraging the discriminator to maximize the log likelihood of the estimations for the correct input source. For task two, the loss function used for $\mathcal{D}$ and $\mathcal{G}$ to maximize the estimation accuracy is formulated as:
\begin{equation}
\begin{aligned}
\mathcal{L}_{\mathcal{D}2},\mathcal{L}_{\mathcal{G}2} = &\max _{\mathcal{D},\mathcal{G}} \mathbb{E}_{x \sim p_{c}(c)}\left[\log y_{2}\right]\\&+\mathbb{E}_{x_ \sim p_{c}(1-c)}\left[\log \left(1-y_{2}\right)\right]
\end{aligned}
\label{eq:lossD2G2}
\end{equation}

Eventually, $\mathcal{D}$ was trained with $\mathcal{L}_{\mathcal{D}1}+\mathcal{L}_{\mathcal{D}2}$, and $\mathcal{G}$ was trained with $\mathcal{L}_{\mathcal{G}1}+\mathcal{L}_{\mathcal{G}2}$. During the training process, we saturated $\mathcal{D}$ before updating the parameters of $\mathcal{G}$, as suggested by \cite{goodfellow2020generative}. More specifically, in each iteration, $\theta_{\mathcal{G}}$ was optimized only when the discriminator was trained to its optimality, i.e., $\theta_{\mathcal{D}}$ completed updates through a stochastic gradient calculation, which complied with the Jensen-Shannon divergence between $p_{g}$ and $p_{r}$. The global minimum of the training criterion was achieved when $p_{g}=p_{r}$, implying that the high-level feature vector synthesized with CT images could approximate the pre-extracted high-level pathological features for the specific class. 
\subsection{Radiological Feature Extraction Module}
The RFEM is another major module of our classification model, which is used to acquire radiographic features from CT images. In this work, we trained the RFEM under the guidance of both the ground truth and the pathological prior knowledge. To employ the synthetic pathological features as the guidance prior, we developed an effective feature fusion framework, where the high-level $512\times1$ pathological feature vector synthesized with CT images and the high-level $512\times1$ radiological feature vector extracted from CT images were concatenated together before being fed to the final FC layer, which was followed by an output layer. Here, we optimized the RFEM with a loss function ${\mathcal{L}_{\text{r}}}$ defined as follows:
\begin{equation}
\begin{aligned}
\operatorname{\mathcal{L}_{\text{r}}}=-\frac{1}{M} \sum_{j} [y_{j} \cdot \log \left(p_{j}\right)+(1-y_{j}) \cdot \log \left(1-p_{j}\right)],
\end{aligned}
\label{eq:radioloss}
\end{equation}
where $j$ refers to the $j^{th}$ sample and $M$ refers to the total number of training samples. $y_{j}$ is the label of the $j^{th}$ sample, and it equals one if the label is the same as LUSC and zero otherwise. $p_{j}$ is the probability that the prediction of the $j^{th}$ sample belongs to LUSC class. It is worth noting that during the training process, the parameters of the well-trained PFSM were fixed. 
\subsection{The Overall Loss Function}
\label{subsec:FF}
The overall objective function $\mathcal{L}_{total}$ of our SGHF-Net classification model is defined as: 
\begin{equation}
\begin{aligned}
{\mathcal{L}_{\text{total}}}=\lambda_{r}\mathcal{L}_{\text{r}}+\lambda_{p}\mathcal{L}_{\text{GAN}}
\end{aligned}
\label{eq:lossG}
\end{equation}
where $\lambda_{r}$ and $\lambda_{p}$ are the tuning parameters used to balance the contributions of the RFEM and the PFSM, respectively; $\mathcal{L}_{\text{GAN}}$ is the representative sign of all objective functions of the GAN.

\begin{table*}[!t]\small
\caption{Clinical information of the multi-center patient cohort.}
\label{tab:databases}
\centering
\renewcommand{\arraystretch}{1.3}
\setlength{\tabcolsep}{1.6mm}{
\begin{tabular}{ccccccccc}
\hline
\multicolumn{9}{c}{\large{Multi-Center Dataset}}  \\ \hline \hline
\multicolumn{3}{c||}{\begin{tabular}[c]{@{}c@{}}Hospital A\\ (Affiliated Dongyang Hospital \\ of Wenzhou Medical University)\end{tabular}}                               
& \multicolumn{3}{c||}{\begin{tabular}[c]{@{}c@{}}Hospital B\\ (Sir Run Run Shaw Hospital, \\Zhejiang University School of Medicine )\end{tabular}}                       
& \multicolumn{3}{c}{\begin{tabular}[c]{@{}c@{}}Hospital C\\ (Cancer Hospital of The University \\of Chinese Academy of Sciences)\end{tabular}}   \\ \hline
\multicolumn{1}{c}{\begin{tabular}[c]{@{}c@{}}Patient\\ (n=191)\end{tabular}}        
& \multicolumn{1}{c}{\begin{tabular}[c]{@{}c@{}}LUAD\\ (n=106)\end{tabular}}        
& \multicolumn{1}{c||}{\begin{tabular}[c]{@{}c@{}}LUSC\\ (n=85)\end{tabular}}       
& \multicolumn{1}{c}{\begin{tabular}[c]{@{}c@{}}Patient\\ (n=388)\end{tabular}}        
& \multicolumn{1}{c}{\begin{tabular}[c]{@{}c@{}}LUAD\\ (n=212)\end{tabular}} 
& \multicolumn{1}{c||}{\begin{tabular}[c]{@{}c@{}}LUSC\\ (n=176)\end{tabular}} 
& \multicolumn{1}{c}{\begin{tabular}[c]{@{}c@{}}Patient\\ (n=250)\end{tabular}}        
& \multicolumn{1}{c}{\begin{tabular}[c]{@{}c@{}}LUAD\\ (n=150)\end{tabular}}        
& \begin{tabular}[c]{@{}c@{}}LUSC\\ (n=100)\end{tabular}       \\ \hline
\multicolumn{1}{c}{Age (year)}      
& \multicolumn{1}{c}{60.43$\pm$11.03} 
& \multicolumn{1}{c||}{66.71$\pm$6.83} 
& \multicolumn{1}{c}{Age (year)}      
& \multicolumn{1}{c}{60.77$\pm$9.39}   
& \multicolumn{1}{c||}{64.13$\pm$7.26}   
& \multicolumn{1}{c}{Age (year)}      
& \multicolumn{1}{c}{61.76$\pm$10.15} 
& 64.52$\pm$8.02 \\ 
\multicolumn{1}{c}{Male}   
& \multicolumn{1}{c}{35}          
& \multicolumn{1}{c||}{85}         
& \multicolumn{1}{c}{Male}   
& \multicolumn{1}{c}{88}   
& \multicolumn{1}{c||}{170}   
& \multicolumn{1}{c}{Male}   
& \multicolumn{1}{c}{61}          
& 97         \\ 
\multicolumn{1}{c}{Female} 
& \multicolumn{1}{c}{71}          
& \multicolumn{1}{c||}{0}          
& \multicolumn{1}{c}{Female} 
& \multicolumn{1}{c}{124}   
& \multicolumn{1}{c||}{6}   
& \multicolumn{1}{c}{Female} 
& \multicolumn{1}{c}{89}          
& 3          \\ \hline
\end{tabular}}
\end{table*}

\subsection{Evaluation Metrics}
\label{subsec:metrics}
We evaluated the performance of our method with three widely-adopted metrics, i.e., the accuracy (ACC), the area under the curve (AUC), and the F1 score. ACC is the ratio of the number of accurate predictions to the number of total model inputs. It is the basic metric for evaluating a classification model and is defined as
\begin{equation}
	ACC=\frac{TP+TN}{TP+TN+FP+FN},
	\label{eq:accuracy}
\end{equation}
where TP, TN, FP and FN represent True Positives, True Negatives, False Positives, and False Negatives, respectively. 

The AUC, which is a helpful addition for evaluating a classification model, refers to the area under the receiver operating characteristic (ROC) curve. The ROC curve is a graphical plot illustrating the true-positive rate (TPR) and false-positive rate (FPR) parameters, which are defined as follows:  
\begin{equation}
	TPR=\frac{TP}{TP+FN}
	\label{eq:TPR}
\end{equation}
\begin{equation}
	FPR=\frac{FP}{FP+TN}
	\label{eq:FPR}
\end{equation}

The F1 score, which is the harmonic mean between precision and recall, is also an efficient criterion for reflecting the performance of a binary classifier. It is defined as
\begin{equation}
	F_{1}=\frac{TP}{TP+1/2(FP+FN)}
	\label{eq:F1}
\end{equation}

All of these metrics have values ranging from 0 to 1, and a higher value indicates better model performance.  
\subsection{Implementation Details}
The architectures of our proposed pathological priors guided classification model were deployed with the PyTorch framework on the Ubuntu 18.04 operating system. The CUDA 10.2 toolkit and CUDA Deep Neural Network (cuDNN) 8.0.2 were utilized to accelerate the model training, and all experiments were conducted on a Tesla V100-SXM2 graphic card. A number of SOTA networks were evaluated as the backbone of our proposed model, including ResNet-18, ResNet-34, ResNet-50, DenseNet-121, DenseNet-169, and the Transformer. All networks were trained from scratch with batch sizes of 2 and 16 for 3D CT images and 2D pathological images, respectively. In addition, a total of 400 epochs were run for each approach, and adaptive moment estimation (Adam) with an initial learning rate of 0.0001 was adopted as the optimizer to update the model parameters.

\section{Dataset}
\label{sec:data}
\begin{figure}
\centering
\includegraphics[width=\linewidth]{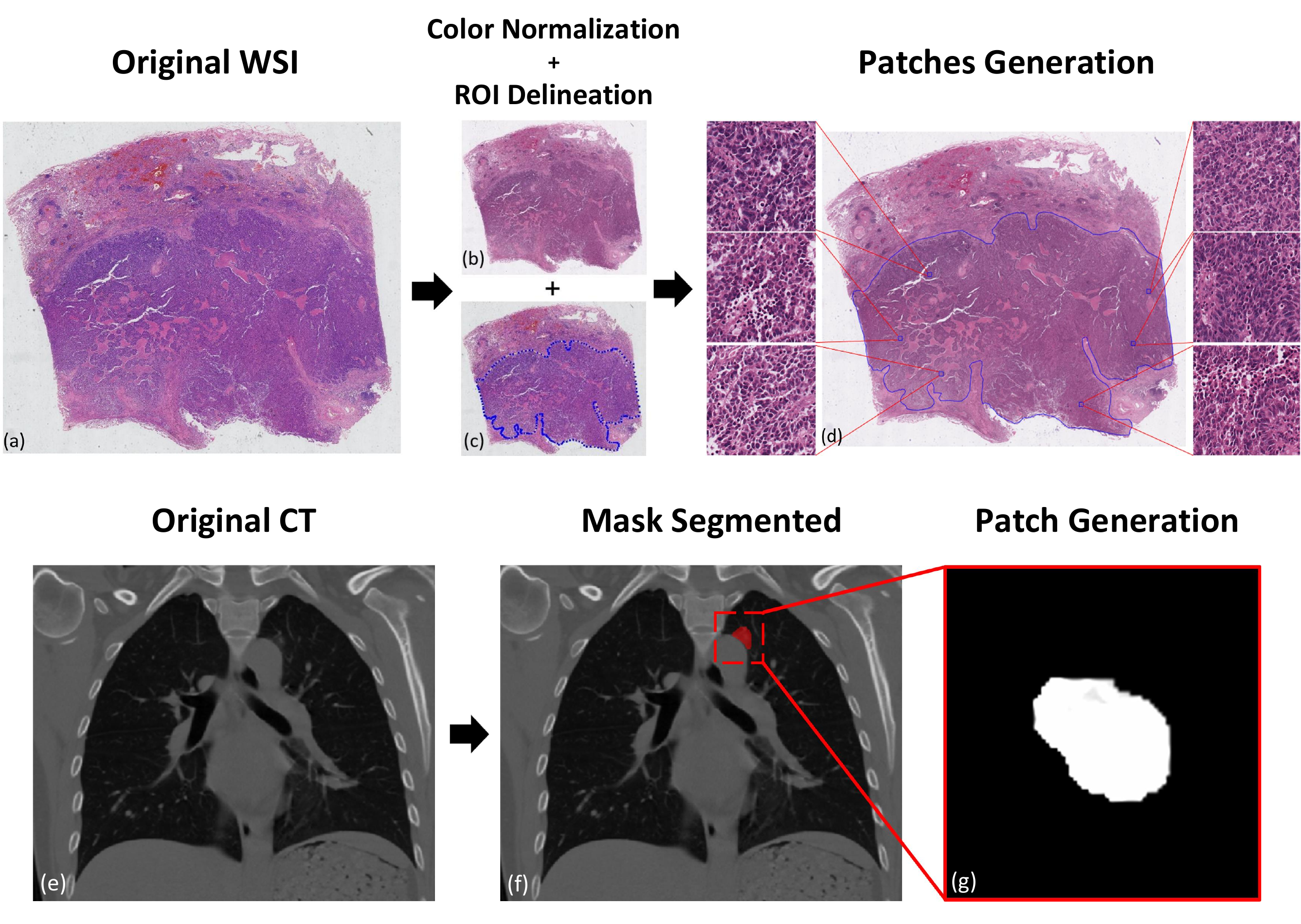}
\caption{The workflow of pathological images and CT images preprocessing: (a) obtaining the original WSI (b) conducting color normalization (c) processing ROI delineation (d) generating patches from WSI with constant size (e) obtaining the original CT image (f) segmenting masks (g) generating patch from CT with constant size.}
\label{fig:preprocessing}
\end{figure}
This study was conducted on a large-scale multi-center dataset formed by three hospitals, the Affiliated Dongyang Hospital of Wenzhou Medical University (hospital A), Sir Run Run Shaw Hospital, Zhejiang University School of Medicine (hospital B), and the Cancer Hospital of The University of Chinese Academy of Sciences (hospital C). Notably, the data retrospectively collected from hospital A, including CT images and their corresponding pathological images, were used for training and testing the proposed model. The data from hospital B and hospital C containing only CT images were applied to evaluate the stability and generalization ability of the well-trained network. Table \ref{tab:databases} demonstrates the demographic characteristics of all patients in this study.

A total of 191 patients from hospital A who had undergone both CT scanning and biopsy/surgical specimen examinations were enrolled in this study. All patients were histopathologically diagnosed with lung cancer with either LUAD or LUSC subtype. To ensure the consistency of the clinical data, we only collected data with time intervals between the pathological examinations and CT examinations that were within two weeks. In addition, the numbers of enrolled patients with final diagnoses of their LUAD/LUSC cancer subtypes from hospital B and hospital C was 388 and 250, respectively. Before training and testing our proposed approach, preprocessing the original pathological images and the multi-center CT images was essential. Brief explanations of the related operations are presented in the following subsections. 

\subsection{Pathological Image Preprocessing}
\label{subsubsec:pathological pre-processing}
The pathological images involved in this work were WSIs, where the sizes of all WSIs were 77460$\pm$14662 (mean$\pm$standard deviation) pixels wide and 59317$\pm$11014 pixels high. With their high resolution, WSIs offered the ultra-fine details of the examined tissue samples, while their excessive file sizes made them a challenge for the proposed learning model to read. Thus, the original WSIs needed to be converted into readable patches with a uniform size. In addition, to ensure the validity of the feature information, color normalization for all WSIs and delineation of the region of interest (ROI) were compulsory before performing patch cropping. Fig. \ref{fig:preprocessing} (a)-(d) illustrates the whole WSI preprocessing procedure.   

The color information contained in tissue slices is critical for pathological histology diagnosis. Although the slices were stained with the same reagents (hematoxylin and eosin, H\&E), the differences among the chemical manufactures or the scanners used for image acquisition may have led to unwanted appearance variations. To avoid the influence of color variations, the technique called structure-preserving color normalization (SPCN) was adopted in this study \cite{vahadane2015structure,anand2019fast}, which properly preserved the stain densities and biological structure.

With respect to ROI delineation, the suspected cancerous regions of all WSIs were delineated by three experienced pathologists on the Automated Slide Analysis Platform (ASAP) \cite{ASAP}. The patches were cropped only within the ROI to ensure that the cancer-related features were contained. After completing the preparations mentioned above, patches with sizes of 560$\times$560 pixels were eventually cropped for each WSI. 
\begin{table*}[!h]
\centering
\caption{The comparisons between the proposed model (marked in red font) and other SOTAs (marked in black font) in terms of ACC(\%), AUC(\%) and F1 score(\%).}
\renewcommand{\arraystretch}{1.5}
\setlength{\tabcolsep}{3mm}{
\begin{tabular}{c c c c c c c}
\hline 
Metrics &  ResNet-18 & ResNet-34 & ResNet-50 & DenseNet-121 & DenseNet-169 & Ours \\ \hline 
ACC 
&	83.13$\pm$3.25 
&	83.30$\pm$4.64 
&	\textbf{83.45$\pm$7.15 }
&	82.95$\pm$2.99 
&	83.13$\pm$4.19 
&87.68$\pm$6.81
\\ 
AUC
& 88.58$\pm$6.11 
& 88.48$\pm$4.18 
& \textbf{89.88$\pm$7.14}
& 88.55$\pm$4.90 
& 88.70$\pm$4.32 
& 91.18$\pm$6.35
\\ 
F1 score        
&	82.03$\pm$5.22 
&	81.95$\pm$3.52 
& \textbf{82.70$\pm$6.65 }
&	82.58$\pm$4.38 
&	82.15$\pm$3.88 
& 86.73$\pm$6.97
\\ \hline \hline
\end{tabular}}
\label{tab:comparisonstudy}
\end{table*}
\subsection{CT Image Preprocessing}
\label{subsubsec:CT pre-processing}
A high-quality set of CT volumes from hospital A was acquired on a PHILIPS Brilliance 64-slice CT system, where the tube voltage and current were 120 kV and 250 mA, respectively. Additionally, CT data from hospital B and hospital C were obtained with a SIEMENS SOMATOM Definition AS 64 CT scanner and a SIEMENS Biograph 16 PET/CT scanner, respectively. To ensure that the malignancies in various patient cases were all captured within the scanning region, the CT data were reconstructed with the same settings: a 0.7 mm in-plane resolution and a 2.0 mm slice thickness. The tumors of the multi-center CT images were segmented by a radiology specialist with the ITK-SNAP 3.8 medical image processing software \cite{yushkevich2006user}. To cover all cancer-related features, the cancer masks were automatically dilated by three voxels during the region delineation procedure. A final volume of interest (VOI) with a constant size of 256$\times$256$\times$128 was cropped for each CT image. Furthermore, the values of all generated patches were normalized between zero and one. The workflow of CT image preprocessing can be viewed in Fig. \ref{fig:preprocessing} (e)-(g).

\section{Experiments}
\label{sec:Results}
To study the proposed classification model SGHF-Net and evaluate its performance, a series of experiments were carried out in this section, consisting of a comparison study with the SOTAs, a ablation study, an impact analysis regarding the number of parameters, a Transformer embedding test, and a multi-center data external validation study. The metrics ACC, AUC, and F1 score were utilized to evaluate the performance of the tested methods. For a better evaluation, the demonstrated results were the average of the five-fold cross-validation.   
\subsection{Comparison with the SOTAs}
To evaluate the feasibility of the proposed SGHF-Net classification model, we conducted a comparison study between the proposed model and five other SOTA classification models, i.e., ResNet-18, ResNet-34, ResNet-50, DenseNet-121, and DenseNet-169. Here, we built the backbone architectures of our model, i.e., the RFEM and the PFSM, by adopting ResNet-18. The SOTA methods were trained from scratch with only the CT images of hospital A as their inputs, while our model was trained with the paired pathological and CT images of hospital A. During the validation phase for each fold, all comparison methods were evaluated on the CT testing dataset.

Ahead of the comparison, the effectiveness of the pathological feature extractor, which exists in our model only during the training phase, was first evaluated. This feature extractor was part of a pre-trained ViT. Taking the pathological images as inputs, this ViT yielded satisfactory results with 93.84\%$\pm$2.23\% ACC and 93.18\%$\pm$1.92\% AUC values for lung cancer subtypes classification, which proved the reliability of its feature extractor.

In Table \ref{tab:comparisonstudy}, the quantitative evaluation results obtained by our model and the five SOTA methods are summarized. The highest value of each metric for all SOTA methods is highlighted in bold. Through the comparisons, we can make the following observations. Overall, our proposed classification model (SGHF-Net) outperformed all of the other classification methods with respect to every metric. For ACC, our approach achieved a result of 87.68\%$\pm$6.81\%, while the highest value among the SOTA methods was provided by ResNet-50 at 83.45\%$\pm$7.15\%. The significant improvements yielded by our model can be confirmed through this 4.23\% difference. For the AUC, the best and the second-best results were 91.18\%$\pm$6.35\% and 89.88\%$\pm$7.14\%, respectively, which were reached by our model and ResNet-50. For the F1 score metric, we realized a 4.03\% improvement, yielding quantitative results of 86.73\%$\pm$6.97\%. These numerical results reveal that the CT-based classification model did improve its performance by incorporating the relevant pathological prior knowledge contained in the pathological images into the RF-based benchmark network. 
\begin{figure*}[!h]
	\centering
	\begin{tabular}{c}
     \includegraphics[width=0.96\textwidth]{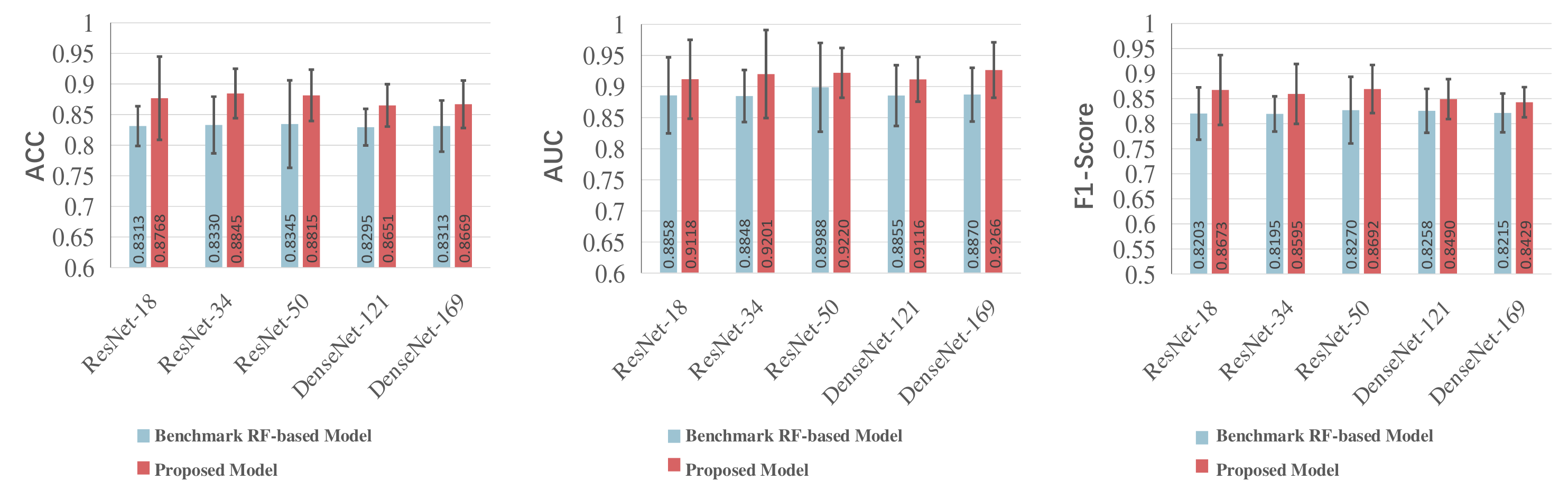} \\  
      (a) \hspace{5.2cm} (b)   \hspace{5.2cm} (c)    \\ 
    \end{tabular}
	\caption{The ablation test of the PFSM in terms of ACC, AUC and F1 score.}
	\label{fig:ablation1}
\end{figure*}
\begin{figure*}[!h]
	\centering
	\begin{tabular}{c}
     \includegraphics[width=0.96\textwidth]{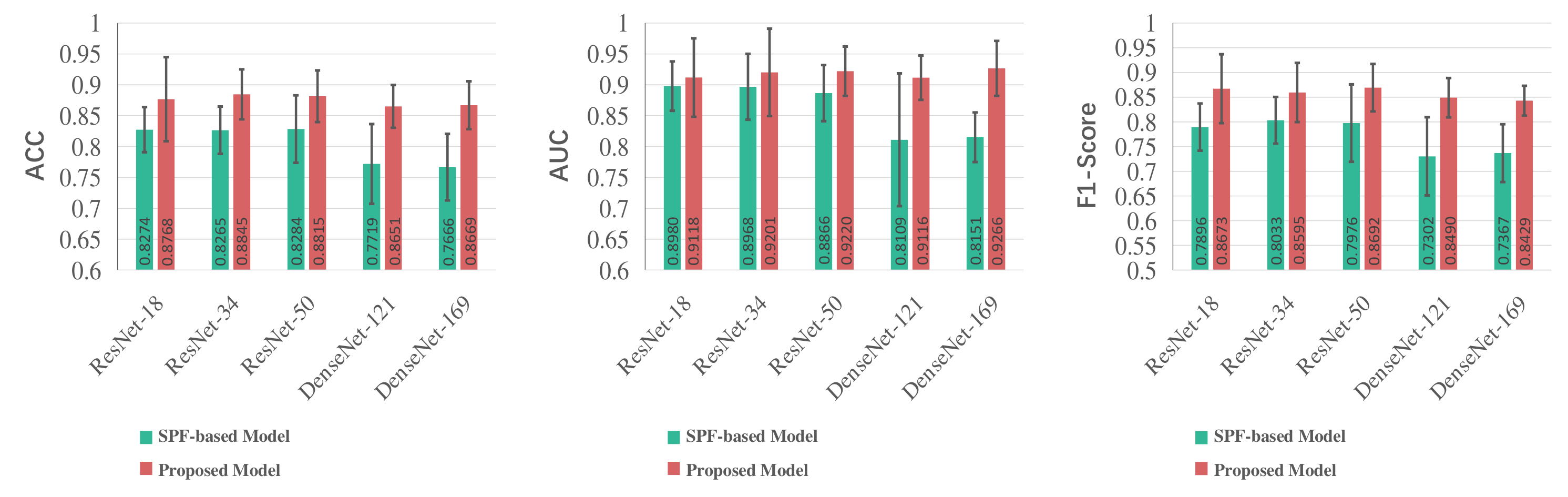} \\  
      (a) \hspace{5.2cm} (b)   \hspace{5.2cm} (c)    \\
    \end{tabular}
	\caption{The ablation test of the RFEM in terms of ACC, AUC and F1 score.}
	\label{fig:ablation2}
\end{figure*}
\begin{figure*}[!h]
	\centering
	\begin{tabular}{ccc}
     \includegraphics[width=0.96\textwidth]{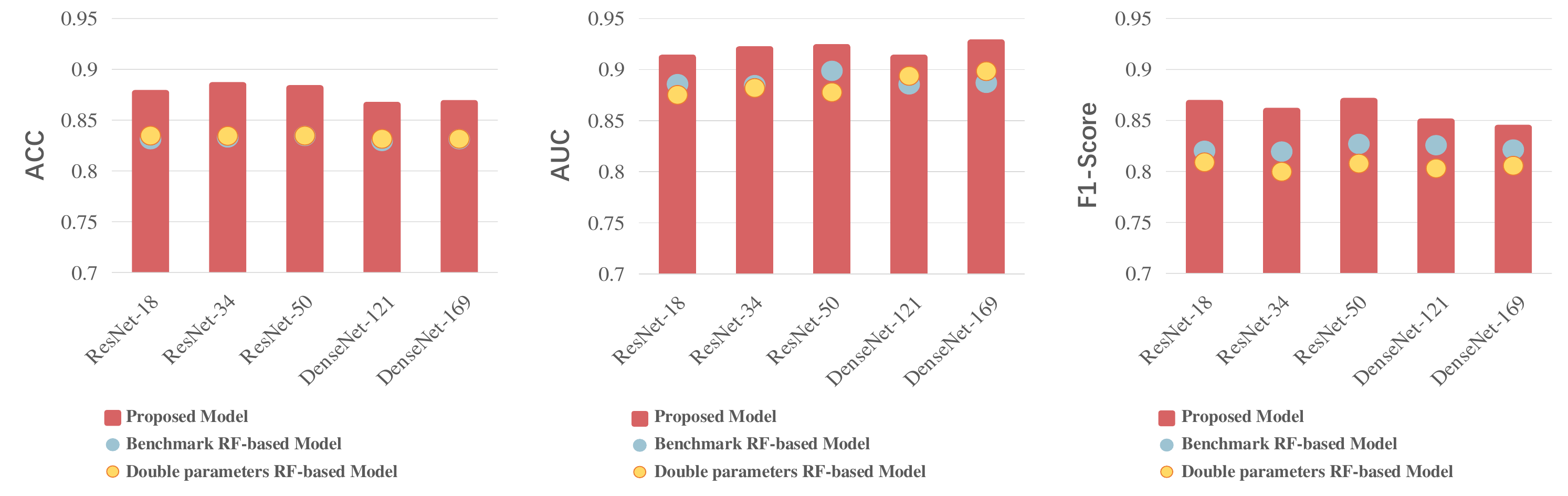} \\
      (a) \hspace{5.2cm} (b)   \hspace{5.2cm} (c)    \\
    \end{tabular}
	\caption{Impact analysis regarding the varying number of parameters in terms of ACC, AUC and F1 score.}
	\label{fig:effectstudy}
\end{figure*}
\subsection{Ablation Study}
\label{subsec:ablation study}
In this work, two key components are contained in the entire design of the proposed model, i.e., the PFSM and the RFEM. To determine the contributions of each module to the achieved performance improvements, an ablation study was carried out.
\subsubsection{The Effectiveness of Synthetic Pathological Priors}
In the above subsection, we demonstrated the superiority of our model through comparisons with other SOTA methods, where we adopted ResNet-18 as the backbone to construct the network. To further investigate the contributions of the PFSM, we also integrated it into the other benchmark models, i.e., ResNet-34, ResNet-50, DenseNet-121, and DenseNet-169, for lung cancer subtypes classification. Correspondingly, the structure of the PFSM was kept consistent with that of the RFEM. To conduct a fair comparison, the weight ratio of the two modules remained 1:1 regardless of which backbone was utilized. 

Illustrated in Fig. \ref{fig:ablation1} (a), (b) and (c), the ACC, AUC, and F1 score were applied for the comparison between the benchmark classification models and the proposed approach with guidance provided by the pathological priors. The effectiveness of the PFSM was prominent. In addition to ResNet-18, taking ResNet-34 as the backbone, the metrics were increased from 83.30\% to 88.45\% for the ACC, 88.48\% to 92.01\% for the AUC, and 81.95\% to 85.95\% for the F1 score. Taking ResNet-50 as the backbone, the ACC, AUC and F1 score increased from 83.45\% to 88.15\%, 89.88\% to 92.20\%, and 82.70\% to 86.92\%, respectively. Taking DenseNet-121 as the backbone, the ACC, AUC and F1 score increased from 82.95\% to 86.51\%, 88.55\% to 91.16\%, and 82.58\% to 84.90\%, respectively. For DenseNet-169, increases from 83.13\% to 86.69\%, 88.70\% to 92.66\%, and 82.15\% to 84.29\% could be seen for the ACC, AUC and F1 score metrics. In general, the integration of synthetic pathological priors into the benchmark RF-based network yielded significant improvements in its classification performance. Therefore, the contributions of the PFSM can be confirmed. 
\subsubsection{The Effectiveness of Radiological Features}
In our design, the final output layer produces classification predications by relying on the hybrid/fused features, which are formed by the concatenation of high-level radiological features and synthetic pathological features. Previous experiments have studied the effects of the synthetic pathological features. Thereafter, to analyze the contributions of the generated radiological features, we built a model that solely relied on the synthetic pathological features for lung cancer subtypes classification, and compared it with the proposed approach. The construction of this model was the same as that of the benchmark radiology-based classification model. To facilitate the following discussion, we name this model as the synthetic-pathological-feature (SPF)-based model. 

In Fig. \ref{fig:ablation2} (a), (b) and (c), the comparisons between the SPF-based model and our proposed model are shown with the same metric set. Likewise, we tested the same five networks as the backbone of the SPF-based model. Comparing the numerical results, it can be observed that the model based only on the SPFs suffered from performance degradation. Taking ResNet-18 as the backbone, the ACC, AUC, and F1 score decreased from 87.68\% to 82.74\%, 91.18\% to 89.80\%, and 86.73\% to 78.96\%, respectively, from our proposed model to the SPF-based model. For ResNet-34, the metrics dropped from 88.45\% to 82.65\% for the ACC, 92.01\% to 89.68\% for the AUC, and 85.95\% to 80.33\% for the F1 score. Taking ResNet-50 as the backbone, the ACC, AUC and F1 score decreased from 88.15\% to 82.84\%, 92.20\% to 88.66\%, and 86.92\% to 79.76\%. Taking DenseNet-121 as the backbone, the decreases from 86.51\% to 77.19\%, 91.16\% to 81.09\%, and 84.90\% to 73.02\% for the ACC, AUC and F1 score metrics, respectively, are evident. For DenseNet-169, the ACC, AUC and F1 score decreased from 86.69\% to 76.66\%, 92.66\% to 81.51\%, and 84.29\% to 73.67\%, respectively. Thus, it implies that RFEM also plays a critical role in the proposed model.  
\begin{table*}[!t]
\centering
\caption{The validation tests of benchmark model (marked in black font) and proposed model (marked in red font) on dataset from hospital B in terms of ACC(\%) and AUC(\%).}
\begin{tabular}{ccccc}
\hline
Metrics & \multicolumn{2}{c}{ACC} & \multicolumn{2}{c}{AUC}  \\ \hline 
Model                    & \multicolumn{1}{c}{\begin{tabular}[c]{@{}c@{}}Benchmark \\ Model\end{tabular}} & \begin{tabular}[c]{@{}c@{}}Proposed \\ Model\end{tabular} 
                         & \multicolumn{1}{c}{\begin{tabular}[c]{@{}c@{}}Benchmark \\ Model\end{tabular}} & \begin{tabular}[c]{@{}c@{}}Proposed \\ Model\end{tabular}    \\ \hline 
ResNet-18	    & \multicolumn{1}{c}{72.64$\pm$2.48} 	& 76.00$\pm$2.3	& \multicolumn{1}{c}{79.17$\pm$3.13}	& 82.07$\pm$3.30	\\
ResNet-34 	    & \multicolumn{1}{c}{69.60$\pm$6.11} 	& 76.05$\pm$2.18	& \multicolumn{1}{c}{78.28$\pm$3.99}	& 83.44$\pm$2.6	     \\
ResNet-50	    & \multicolumn{1}{c}{72.88$\pm$2.71} 	& 76.16$\pm$3.37	& \multicolumn{1}{c}{79.23$\pm$3.08}	& 83.30$\pm$3.36	\\
DenseNet-121  	& \multicolumn{1}{c}{72.60$\pm$2.65} 	& 75.20$\pm$2.04	& \multicolumn{1}{c}{77.82$\pm$5.18}	& 81.20$\pm$2.08     \\
DenseNet-169	& \multicolumn{1}{c}{73.04$\pm$1.54}	& 75.92$\pm$2.56	& \multicolumn{1}{c}{79.68$\pm$2.37}	& 82.12$\pm$1.44	\\ 
ViT & \multicolumn{1}{c}{73.77$\pm$3.13}	& 76.89$\pm$3.91	& \multicolumn{1}{c}{80.07$\pm$4.33}	& 84.29$\pm$4.12	\\ \hline
\end{tabular}
\label{tab:validationB}
\end{table*}
\begin{table*}[!t]
\centering
\caption{The validation tests of benchmark model (marked in black font) and proposed model (marked in red font) on dataset from hospital C in terms of ACC(\%) and AUC(\%).}
\begin{tabular}{ccccc}
\hline
Metrics & \multicolumn{2}{c}{ACC} & \multicolumn{2}{c}{AUC}  \\ \hline 
Model                    & \multicolumn{1}{c}{\begin{tabular}[c]{@{}c@{}}Benchmark \\ Model\end{tabular}} & \begin{tabular}[c]{@{}c@{}}Proposed \\ Model\end{tabular} 
                         & \multicolumn{1}{c}{\begin{tabular}[c]{@{}c@{}}Benchmark \\ Model\end{tabular}} & \begin{tabular}[c]{@{}c@{}}Proposed \\ Model\end{tabular}    \\ \hline 
ResNet-18	    & \multicolumn{1}{c}{71.11$\pm$4.32}	& 74.89$\pm$2.12 	& \multicolumn{1}{c}{75.24$\pm$2.33}	& 79.15$\pm$4.49 	\\
ResNet-34 	    & \multicolumn{1}{c}{70.44$\pm$5.90}	& 75.58$\pm$2.9	& \multicolumn{1}{c}{75.08$\pm$2.90}	& 79.72$\pm$4.76	\\
ResNet-50	    & \multicolumn{1}{c}{72.08$\pm$2.88}	& 74.62$\pm$1.61	& \multicolumn{1}{c}{76.02$\pm$2.92}	& 78.91$\pm$4.84	\\
DenseNet-121  	& \multicolumn{1}{c}{72.27$\pm$2.42}	& 74.29$\pm$3.6	& \multicolumn{1}{c}{74.01$\pm$4.32}	& 77.60$\pm$2.49	\\
DenseNet-169	& \multicolumn{1}{c}{71.21$\pm$3.55}	& 73.67$\pm$2.72	& \multicolumn{1}{c}{75.04$\pm$3.55}	& 76.50$\pm$4.40	\\ 
ViT & \multicolumn{1}{c}{72.76$\pm$2.79}	& 74.97$\pm$3.07	& \multicolumn{1}{c}{76.47$\pm$3.54}	& 79.69$\pm$4.31	\\ \hline
\end{tabular}
\label{tab:validationC}
\end{table*}
\subsection{Impact Analysis Regarding the Number of Parameters}
Compared with the benchmark RF-based classification model, our approach exhibits significant accuracy improvements. However, it is worth noting that the number of parameters in our model is twice as much as that in the benchmark model. To study the impact of the number of parameters on the model's performance, we trained a model that had the same structure as the proposed method while replacing its PFSM with another RFEM. To be more specific, this model, namely, the double parameters RF-based model, was made of two paralleled RFEM and a FC layer, under the same feature fusion framework as that of the proposed method. As shown in Fig. \ref{fig:effectstudy}, the corresponding ACC, AUC and F1 score values were compared among the benchmark RF-based model, the double parameters RF-based model, and our proposed model. When adopting ResNet-18, ResNet-34, ResNet-50, DenseNet-121, and DenseNet-169 as the backbone, the numbers of parameters in the benchmark model were 10.80\textbf{m}, 18.38\textbf{m}, 26.88\textbf{m}, 1.96\textbf{m}, and 3.01\textbf{m}, while that of the benchmark model with twice as many parameters and our model were 21.60\textbf{m}, 36.76\textbf{m}, 53.76\textbf{m}, 3.92\textbf{m}, and 6.02\textbf{m}, where \textbf{m} refers to million. Comparing the metrics in Fig. \ref{fig:effectstudy}, it can be concluded that doubling the number of parameters did not dominate the optimization of the model.  
\subsection{Transformer Embedding Test}
In addition to the above-mentioned SOTA methods, we further extended our proposed model by replacing the ResNet backbone with a powerful Transformer network. Similarly, comparison studies between the benchmark RF-based model and our proposed model were conducted. For the benchmark model that was equipped with the Transformer, the ACC, AUC and F1 score were 83.56\%$\pm$5.84\%, 90.06\%$\pm$5.99\% and 81.42\%$\pm$7.29\%, respectively, which were slightly better than the best performance of the above tested SOTA methods but far from the results of our approach with ResNet backbone. Furthermore, by adapting the network of our proposed method with the Transformer, 88.04\%$\pm$5.29\%, 92.48\%$\pm$6.10\%, and 85.53\%$\pm$7.05\% ACC, AUC, and F1 score values could be obtained, demonstrating remarkable improvements. These results confirm the great value of the additional pathological priors information in accurately performing medical classification tasks.  
\subsection{External Validation}
\label{subsec:external validation}
This study was conducted on a large-scale multi-center dataset, which was formed with data acquired from three hospitals. With the training dataset from hospital A, the advancements of our proposed model over the benchmarks were shown through a number of experiments. Moreover, it can be claimed that the achieved improvements are attributed to the PFSM. In this subsection, we applied the well-trained model on the CT dataset from hospitals B and C to validate its generalization ability and robustness. In Table \ref{tab:validationB} and \ref{tab:validationC}, the external performance validation results obtained for the data acquired from these two hospitals are presented, showing the comparisons between the benchmark RF-based model and our proposed model. Judging from the results, though it has to be admitted that both the benchmark model and our model performed worse on the data from hospitals B and C than those from hospital A, these models still provided acceptable ACCs and AUCs. More importantly, on the same dataset, the proposed classification model SGHF-Net attained superior performance to that of the benchmark network, achieving both ACC and AUC improvements.

\section{Discussion}
\label{sec:Discussion}
The key challenge encountered when using DL algorithms in clinical applications is to achieve high and stable diagnostic accuracy. Studies have achieved improved model performance by designing more complex structures or increasing the number of utilized parameters. In this study, we proposed an original strategy and framework by applying the multi-modality hybrid features to enhance the network training process. This model was trained with paired pathological images and CT images, while it relied only on CT images during the validation and testing phases. Hence, the novelty of the proposed model lies in its ability to self-generate hybrid features with single-modality inputs. From the results in Table \ref{tab:comparisonstudy}, \ref{tab:validationB} and \ref{tab:validationC}, it can be seen that our framework yielded obviously higher accuracies than other SOTA methods on both the local-center validation dataset and multi-center validation dataset.

In section \ref{subsec:ablation study}, we further proved the effectiveness of the PFSM and RFEM through an ablation study. By separately removing the PFSM and RFEM from the complete network, we confirmed the significance of both parts in the whole model. Regarding the PFSM, the synthetic pathological features not only enhanced the specificity of the pathology-related information, but also suppressed the interference caused by irrelevant information in the original CT images. Moreover, as demonstrated in Fig. \ref{fig:ablation1}, the pathological priors guided strategy could be easily adopted by other classification models to achieve improved performance. In addition to the PFSM, the RFEM also played an important role in the entire framework, as represented in Fig. \ref{fig:ablation2}. From a comprehensive perspective, the radiological features obtained by RFEM provided constraints to reduce network degeneration of PFSM. Therefore, these two modules are complementary to each other for a more accurate cancer subtypes classification.

During the training procedure of the PFSM, we designed an improved strategy for extracting pathological features from WSIs. Considering the limitation of the GPU memory in this work, it was impossible to directly extract pathological features from the entire WSIs with large image sizes. We chose to apply the ViT network to approximate the feature extraction design. Patches with constant image sizes were randomly collected from the original WSIs and embedded into tokens. Therefore, the pathological feature of one WSI could be obtained from all the patches based on their attention values. The other feature strategies involved converting patches into features and then selecting the most suitable features for the original WSIs, which was proven to be less efficient for this task than the ViT strategy. However, using patches extracted from WSIs with a constant size, this approach still has limitations in representing the global information contained in WSIs. In future research, a more comprehensive multi-scale feature extraction strategy may be further developed to acquire more abundant pathological information.

In section \ref{subsec:external validation}, external datasets were applied to validate the multi-center performance of our network. According to Table \ref{tab:validationB} and \ref{tab:validationC}, it could be concluded that our proposed classification model was superior to other SOTA methods. However, compared with their validation results for hospital A, all models exhibited non-negligible degradation on the multi-center datasets, whether they included our pathological priors guided strategy or not. Under multi-center circumstances, differences of CT scanners or varying operation habits of radiologists may result in inevitable deviations in the image resolutions or CT values for the same lesion, which is a major challenge with respect to the model's generalization ability. Comparing the results obtained on different datasets, our framework had no significant advantage in dealing with multi-center effects. The introduction of clinical information in future studies may be one of the solutions to this multi-center problem.

In future work, we will further explore the feasibility of implementing the proposed approach in more complex applications, e.g. multi-focal cancer/combined cancer diagnosis with CT images, cancer invasion degree assessment with CT images, and other cancer-related clinical problems. In addition, we can also extend this framework to wider areas and solve more multi-modal tasks.

\section{Conclusion}
\label{sec:Conclusion}
In this paper, we proposed a novel deep learning network, SGHF-Net, for accurately classifying lung cancer subtypes on CT images. The proposed model was constructed under a feature fusion framework with two key components, a pathological feature synthetic module (PFSM) and a radiological feature extraction module (RFEM). These two modules assisted the network in acquiring both high-level pathological features and high-level radiological features from CT images alone, where the pathological features contained information about the corresponding pathological images. Following the feature fusion framework, the hybrid features covering both pathological priors and radiographic information could then be obtained by integrating the two modules together. Through a series of strategy comparisons, we demonstrated the superiority of the proposed framework and confirmed that the complementarity of synthetic pathological priors (i.e., pathological features) could significantly improve the accuracy of CT-based lung cancer subtypes classification approaches. Additionally, the proposed pathological priors guided strategy has considerable potential to be extended to more complicated applications. 

\section*{Acknowledgments}
This work was supported by the National Natural Science Foundation of China (No. 62001425) and the Key Research and Development Program of Zhejiang Province (No. 2021C03029).

\bibliographystyle{unsrt}  
\bibliography{refs}
\end{document}